\documentclass[preprint,showpacs,preprintnumbers,amsmath,amssymb]{revtex4}

\usepackage{graphicx}
\usepackage{epsfig}		
\usepackage{dcolumn}
\usepackage{bm}

\def\0{\mbox{\tiny $0$}}
\def\1{\mbox{\tiny $1$}}
\def\2{\mbox{\tiny $2$}}
\def\3{\mbox{\tiny $3$}}
\def\4{\mbox{\tiny $4$}}
\def\5{\mbox{\tiny $5$}}
\def\6{\mbox{\tiny $6$}}
\def\7{\mbox{\tiny $7$}}
\def\8{\mbox{\tiny $8$}}
\def\9{\mbox{\tiny $9$}}

\def\f14{\mbox{\tiny $\frac{1}{4}$}}
\def\infm{\mbox{\tiny $-\infty$}}
\def\infp{\mbox{\tiny $+\infty$}}

\def\ii{\mbox{\tiny $i$}}
\def\z{\mbox{\tiny $z$}}

\def\j{\mbox{\tiny $j$}}

\def\mi{\mbox{\tiny $-$}}

\def\bb#1{\mbox{\footnotesize $(#1)$}}

\begin{document}


\title{Second-order corrections to neutrino two-flavor oscillation parameters in the wave packet approach}

\author{A. E. Bernardini}
\email{alexeb@ifi.unicamp.br}
\author{M. M. Guzzo}
\email{guzzo@ifi.unicamp.br}
\author{F. R. Torres}
\email{ftorres@ifi.unicamp.br}
\affiliation{Instituto de F\'{\i}sica Gleb Wataghin, UNICAMP,\\
PO Box 6165, 13083-970, Campinas, SP, Brasil.}

\date{\today}

\begin{abstract}
We report about an analytic study involving the {\em intermediate} wave packet formalism for quantifying the physically relevant information which appear in the neutrino two-flavor conversion formula and help us to obtain more precise limits and ranges for neutrino flavor oscillation.
By following the sequence of analytic approximations where we assume a strictly peaked momentum distribution and consider the second-order corrections in a power series expansion of the energy, we point out a {\em residual} time-dependent phase which, coupled with the {\em spreading/slippage} effects, can subtly modify the neutrino oscillation parameters and limits.
Such second-order effects are usually ignored in the relativistic wave packet treatment, but they present an evident dependence on the propagation regime so that some small modifications to the oscillation pattern, even in the ultra-relativistic limit, can be quantified.
These modifications are implemented in the confront with the neutrino oscillation parameter range (mass-squared difference $\Delta m^{\2}$ and the mixing-angle $\theta$) where we assume the same wave packet parameters previously noticed in the literature in a kind of {\em toy model} for some reactor experiments. 
Generically speaking, our analysis parallels the recent experimental purposes which concern with higher precision parameter measurements.
To summarize, we show that the effectiveness of a more accurate determination of $\Delta m^{\2}$ and $\theta$ depends on the wave packet width $a$ and on the averaged propagating energy flux $\bar{E}$ which still correspond to open variables for some classes of experiments.
\end{abstract}

\maketitle

\section{Introduction}

In the last years, the quantum mechanics of neutrino oscillations \cite{Zub98,Alb03,Vog04} has experienced much progress on the theoretical front \cite{Beu03}, not only in phenomenological pursuit of a more refined flavor conversion formula \cite{Giu98,Zra98,Ber05}, but also in efforts to give the theory a formal structure within quantum field formalism \cite{Bla95,Giu02B,Bla03}.
From the point of view of a first quantized theory and in the context of vacuum oscillations, as a first analysis, the probability that neutrinos originally created as a $\mbox{\boldmath$\nu_\alpha$}$ flavor-eigenstate with averaged energy $\bar{E}$ oscillates into a $\mbox{\boldmath$\nu_\beta$}$ flavor-eigenstate over a distance $L$ is given by
\small\begin{equation}
P(\mbox{\boldmath$\nu_\alpha$}\rightarrow\mbox{\boldmath$\nu_\beta$};L)=
\sin^{\2}{(2\theta)}\,\sin^{\2}{\left[\frac{\Delta m^{\2}_{\ii\j}\, L}{4 \bar{E}}\right]}, ~~(c,\hbar = 1)
\label{00}
\end{equation}\normalsize
where we have assumed that the main aspects of the oscillation phenomena can be understood by studying the simple-minded two-flavor problem constructed in terms of the $\mbox{\boldmath$\nu_{\1}$}\bb{x}$ and $\mbox{\boldmath$\nu_{\2}$}\bb{x}$ plane-wave mass-eigenstates with the two-flavor mixing angle represented by $\theta$ and the mass-squared difference given by $\Delta m^{\2}_{\ii\j}$.
As appointed by the D. Groom PDG review \cite{Gro04}, although this equation is frequently quoted and used in Monte Carlo calculations, the wave function is badly behaved for reasons larger than about one, where it oscillates more and more rapidly in the interval between $\sin{(2\theta)} = 0$ and $\sin{(2\theta)} = \langle P \rangle$ as the argument $\frac{\Delta m^{\2}_{\ii\j}\, L}{4 \bar{E}}$ increases.
Moreover, it is difficult to relate this function to the exclusion curves described in the literature \cite{Gro04}.    
  
In fact, the {\em intermediate} wave packet (WP) approach \cite{Kay81} eliminates the most controversial
points rising up with the {\em standard} plane-wave (PW) formalism \cite{Kay89,Kay04}
since wave packets describing propagating mass-eigenstates guarantee the existence of a coherence length \cite{Kay81},
avoid the ambiguous approximations in the PW derivation of the phase difference \cite{DeL04} and, under particular conditions of
minimal {\em slippage} between the mass-eigenstate wave packets, recovers the oscillation probability given by the {\em standard} PW
treatment \footnote{Strictly speaking, the intermediate WP formalism can also be refuted
since oscillating neutrinos are neither prepared nor observed \cite{Beu03} in this case.
Some authors suggest the calculation of a transition probability between
the observable particles involved in the production and detection
process in the so-called {\em external}
WP approach \cite{Giu02B,Beu03,Ric93}: the oscillating
particle, described as an internal line of a Feynman diagram by a
relativistic mixed scalar propagator, propagates between the
source and target (external) particles represented by wave
packets. It can be demonstrated \cite{Beu03}, however, that the overlap function of
the incoming and outgoing wave packets
in the external WP model is mathematically equivalent
to the wave function of the propagating mass-eigenstate in the
intermediate WP formalism.}.
Enclosed by a restrict motivation, we avoid a more extensive discussion about several
controversial aspects concerning with the
intermediate WP formalism (see for instance \cite{Beu03,Fie03} which work on the {\em external} WP framework) and,
quite generally, we follow an analytical approach where the mass-eigenstate
time evolution does not concern with the WP limitations.
As preliminary investigation we consider {\em gaussian} wave packets \cite{Kay81,Giu98,Giu02B}, as we know, the unique which enable
us to analytically quantify the first and the second-order corrections to the flavor conversion formulas.
By means of the second-order terms we can compute the effects of
an extra time-dependent phase which is added to the {\em standard} oscillation term
$\frac{\Delta m^{\2} \,t}{2\, \bar{E}}$ \cite{Kay04} and modifies the
oscillating character of the propagating particles.
Generically speaking, we avoid more sophisticated methods as field theoretical prescriptions \cite{Beu03,Gri99} in detriment to a clearer 
treatment which commonly simplifies the understanding of physical aspects going with the oscillation
phenomena.

Reporting to such an analytical construction \cite{Ber04},
we look for improving the procedure for obtaining the two-flavor oscillation parameter exclusion region boundary
for a generic class of oscillation experiments.
In fact, precise determination of the oscillation parameters and search for non-standard physics such as a small
admixture of a sterile component in the solar neutrino-flux are still of interest.
In addition, from the experimental point of view, to determine $\Delta m^{\2}$ more precisely, further KamLAND exposure to
the reactor neutrinos are becoming most powerful \cite{Bah03,Ban03} at the same time that more precise neutral current measurements
by SNO can contribute in reducing the uncertainty of the mixing angle \cite{Ban03,Ban03B}.
By knowing the emergence of neutrino physics is fueled by such a recent growth in quality and quantity
of experimental data, our main purpose is discussing
how much the determination of mixing parameters and mass-differences can be improved in the context of the wave packet phenomenological analysis and,
eventually, suggest a perspective of improvement on bounds on $\theta_{\1\3}$ 
before experiments designed specifically for this parameter start.
The first step of our study, which is presented in section II,
concerns with the analytical derivation of a flavor conversion formula 
where a {\em gaussian} momentum distribution and
a power series expansion of the energy up to the second-order terms
are introduced for obtaining analytically integrable probabilities.
Adopting a strictly peaked momentum distribution
to construct each mass-eigenstate wave packet allows us to analytically quantify 
these rising-up second-order corrections.
Firstly, we reproduce the localization effects described in terms of the well-known wave packet {\em spreading}
and of the {\em slippage} between the wave packets, which leads to the decoherence between the propagating mass-eigenstates.
In particular, it is well-established that the decoherence effect has a correspondence with an oscillation {\em damping} (exponential) factor \cite{Kay81,Kie96},
thus we just report about such effects to show that both of them
can be quantified, up to second-order corrections, in non-relativistic (NR) and ultra-relativistic (UR) propagation regimes.
In parallel, we also recover an additional time-dependent phase which changes the {\em standard}
oscillating character of the flavor conversion formula.
However, the main contribution of our study, which is presented in section III, concerns with the understanding of the physical aspects carried by neutrino two-flavor oscillation parameters and limits.
In particular, we suggest the existence of more precise corrections in the procedure \cite{Gro04} where the wave packet characteristics
are used for describing the mass-squared difference $\Delta m^{\2}$ and the mixing-angle $\theta$ which are constrained
by the experimental {\em input} parameters.
In section IV, 
by the time we assume the same {\em testing} wave packet parameters previously introduced in the literature \cite{Gro04}
(for some reactor experiments), we can establish a phenomenological comparison which
allows us to confront the energy second-order corrections introduced
in the previous sections with the predicted PW and WP with first-order correction results.
We also observe the possibilities for extending the analysis to other production-{\em type} neutrinos (solar and supernova) . 
Finally, we draw our conclusions in section V by emphasizing that the effectiveness of a more accurate determination of $\Delta m^{\2}$ and $\theta$ 
can eventually depend on the wave packet width $a$ and on the averaged propagating energy flux $\bar{E}$.

\section{Wave packets with second-order corrections}

The time evolution of flavor wave packets can be described by the state vector
\small\begin{eqnarray}
\Phi(z,t) &=& \phi_{\1}(z,t)\cos{\theta}\,\mbox{\boldmath$\nu_{\1}$} + \phi_{\2}(z,t)\sin{\theta}\,\mbox{\boldmath$\nu_{\2}$}\nonumber\\
          &=& \left[\phi_{\1}(z,t)\cos^{\2}{\theta} + \phi_{\2}(z,t)\sin^{\2}{\theta}\right]\,\mbox{\boldmath$\nu_\alpha$}+
		  \left[\phi_{\2}(z,t) - \phi_{\1}(z,t)\right]\cos{\theta}\sin{\theta}\,\mbox{\boldmath$\nu_\beta$}\nonumber\\
          &=& \phi_{\alpha}(z,t;\theta)\,\mbox{\boldmath$\nu_\alpha$} + \phi_{\beta}(z,t;\theta)\,\mbox{\boldmath$\nu_\beta$},
\label{0}
\end{eqnarray}\normalsize
with flavor and mass-eigenstate indices as previously defined.
The probability of finding a flavor state $\mbox{\boldmath$\nu_\beta$}$ at
the instant $t$ is equal to the integrated squared modulus of the
$\mbox{\boldmath$\nu_\beta$}$ coefficient
\small\begin{equation}
P(\mbox{\boldmath$\nu_\alpha$}\rightarrow\mbox{\boldmath$\nu_\beta$};t)=
\int_{_{\infm}}^{^{\infp}}\hspace{-0.5cm}dz \,\left|\phi_{\beta}(z,t;\theta)\right|^{\2}
= 
\frac{\sin^{\2}{(2\theta)}}{2}\left\{\, 1 - \mbox{\sc Int}(t) \, \right\},
\label{1}
\end{equation}\normalsize
where $\mbox{\sc Int}(t)$ represents the mass-eigenstate interference term given by
\small\begin{equation}
\mbox{\sc Int}(t) = Re 
 \left[\, \int_{_{\infm}}^{^{\infp}}\hspace{-0.5cm}dz
\,\phi^{\dagger}_{\1}(z,t) \, \phi_{\2}(z,t) \, \right].\,
\label{2}
\end{equation}\normalsize
Let us consider mass-eigenstate wave packets given by
\small\begin{equation}
\phi_{\ii}(z,0) = \left(\frac{2}{\pi a^{\2}}\right)^{ \frac{1}{4}} \exp{\left[- \frac{z^{\2}}{a^{\2}}\right]} \exp{[i p_{\ii} \, z]},
\label{3}
\end{equation}\normalsize
at time $t = 0$, where $i = 1,\, 2$.
The wave functions which describe their time evolution are
\small\begin{eqnarray}
\phi_{\ii}(z,t) =
\int_{_{\infm}}^{^{\infp}}\hspace{-0.25cm}\frac{dp_{\z}}{2 \pi} \,
\varphi(p_{\z} \mi p_{\ii}) \exp{\left[-i\,E^{(\ii)}_{p_{\z}}\,t +i \, p_{\z}
\,z\right]},
\label{4}
\end{eqnarray}\normalsize
where
$E^{(\ii)}_{p_{\z}} = \left(p_{\z}^{\2} + m_{\ii}^{\2}\right)^{ \frac{1}{2}}$
and
$\varphi(p_{\z} \mi p_{\ii}) =  \left(2 \pi a^{\2} \right)^{ \frac{1}{4}} \exp{\left[- \frac{(p_{\z} \mi p_{\ii})^{\2}\,a^{\2}}{4}\right]}.$
In order to obtain the oscillation probability, we can calculate the interference term $\mbox{\sc Int}(t)$
by solving the following integral
\small\begin{eqnarray}
\lefteqn{\int_{_{\infm}}^{^{\infp}}\hspace{-0.25cm}\frac{dp_z}{2 \pi} \,  \varphi(p_z \mi p_{ 1}) \varphi(p_z \mi p_{ 2})
\exp{[-i \, \Delta E_{p_z} \, t]} =}\nonumber\\ 
&& \exp{\left[ \frac{\mi(a \, \Delta{p})^{\2}}{8}\right]}
\int_{_{\infm}}^{^{\infp}}\hspace{-0.25cm}\frac{dp_z}{2 \pi}  \, \varphi^{\2}(p_z \mi p_{\0})\exp{[-i \, \Delta E_{p_z} \, t]},~~~ \label{6}
\end{eqnarray}\normalsize
where we have changed the $z$-integration into a $p_{\z}$-integration
and introduced the quantities $\Delta p = p_{ \1} \mi p_{ \2}
,\,\, p_{\0} = \frac{1}{2}(p_{ \1} + p_{ \2})$ and $\Delta E_{p_{\z}} = E^{(\1)}_{p_{\z}} \mi E^{(\2)}_{p_{\z}}$.
The oscillation term is
bounded by the exponential function of $a \, \Delta p$ at
any instant of time. Under this condition we could never observe a
{\em pure} flavor-eigenstate. Besides, oscillations are
considerably suppressed if $a \, \Delta p > 1$. A necessary
condition to observe oscillations is that $a \, \Delta p \ll 1$.
This constraint can also be expressed by $\delta p \gg \Delta p$
where $\delta p$ is the momentum uncertainty of the particle. The
overlap between the momentum distributions is indeed relevant only
for $\delta p \gg \Delta p$. 
Strictly speaking, we are assuming that the oscillation length 
($\pi\frac{4 \bar{E}}{\Delta m^{\2}_{\ii\j}}$) is sufficiently larger
than the wave packet width which simply says that
the wave packet must not extend as wide as the oscillation
length, otherwise the oscillations are washed out \cite{Kay81,Gri96,Gri99}.

Turning back to the Eq.~(\ref{6}), without loss of
generality, we can assume
\small\begin{equation}
\mbox{\sc Int}(t) = Re \left\{\int_{_{\infm}}^{^{\infp}}\hspace{-0.25cm}\frac{dp_{\z}}{2
\pi}
 \, \varphi^{\2}(p_{\z} \mi p_{\0})\exp{[-i \, \Delta E_{p_{\z}} \, t]}\right\}
\label{9}.
\end{equation}\normalsize

In the literature, this equation is often obtained by assuming two
mass-eigenstate wave packets described by the same momentum
distribution centered around the average momentum
$\bar{p} = p_{\0}$. This simplifying hypothesis  also guarantees
{\em instantaneous} creation of a {\em pure} flavor
eigenstate {\boldmath$\nu_\alpha$} at $t = 0$ \cite{DeL04}. In fact, for $
\phi_{\1}(z,0)=\phi_{\2}(z,0)$ we get from Eq.~(\ref{0})
\small\begin{equation}
\phi_{\alpha}(z,0,\theta) = \left(\frac{2}{\pi
a^{\2}}\right)^{\frac{1}{4}} \exp{\left[- \frac{z^{\2}}{a^{\2}}\right]}
\exp{[i  p_{\0} \,z]}
\nonumber\end{equation}\normalsize
and
$\phi_{\beta}(z,0,\theta) =0$.
In order to obtain an expression for $\phi_{\ii}(z,t)$ by analytically solving the integral in Eq.~(\ref{4}) we firstly rewrite the energy $E^{(\ii)}_{p_{\z}}$ as
\small\begin{equation}
E^{(\ii)}_{p_{\z}} = E_{\ii} \left[1 + \frac{ p_{\z}^{\2} \mi p_{\0}^{\2}}{E_{\ii}^{\2}}\right]^{ \frac{1}{2}} = E_{\ii} \left[1 + \sigma_{\ii} \left(\sigma_{\ii} + 2 \mbox{v}_{\ii}\right)\right]^{ \frac{1}{2}},
\label{11}
\end{equation}\normalsize
where $E_{\ii} = (m_{\ii}^{\2} + p_{\0}^{\2})^{ \frac{1}{2}}$, $\mbox{v}_{\ii} = \frac{ p_{\0}}{E_{\ii}}$ and 
$\sigma_{\ii} = \frac{ p_{\z} \mi  p_{\0}}{E_{\ii}}$.

We are attentive to the fact that the assumption of wave packets with {\em gaussian} shape
can introduce some (not quantifiable) limitations to the interpretation of the following results.
However, in this kind of analysis and, for instance, in non-relativistic quantum mechanics,
free-propagating {\em gaussian} wave packets are frequently assumed because
the calculations can be carried out exactly for these particular functions and, consequently,
the main physical aspects can be easily interpreted from the final analytically obtained expressions.
The reason lies in the fact that the frequency components
of the mass-eigenstate wave packets,
$E^{(\ii)}_{p_{\z}}= p_{\z}^{\2}/2 m_{\ii}$, modify the momentum
distribution into ``generalized'' {\em gaussian}, easily integrated by
well-known methods of analysis. The term $ p_{\z}^{\2}$ in
$E^{(\ii)}_{p_{\z}}$ is then responsible for the variation in time
of the width of the mass-eigenstate wave packets, the so-called
{\em spreading} phenomenon. In relativistic quantum mechanics the
frequency components of the mass-eigenstate wave packets,
$E^{(\ii)}_{p_{\z}}=\sqrt{ p_{\z}^{\, \2} + m_{\ii}^{\2} }$, do not
permit  an immediate analytic integration. This difficulty,
however, may be remedied by assuming a sharply peaked
momentum distribution, i. e. $(a \, E_{\ii})^{-1}\sim\sigma_{\ii} \ll 1$.
Meanwhile, the integral in Eq.~(\ref{4}) can be {\em
analytically} solved only if we consider terms up to order
$\sigma_{\ii}^2$ in the series expansion.
In this case, we can conveniently
truncate the power series
\small\begin{eqnarray}
E^{(\ii)}_{p_{\z}} & = & E_{\ii} \left[1 + \sigma_{\ii} \mbox{v}_{\ii}  + \frac{\sigma_{\ii}^{\2}}{2}\left(1 - \mbox{v}_{\ii}^{\2} \right)\right] + \mathcal{O}(\sigma_{\ii}^{\3})
\nonumber\\  &
 \approx &
  E_{\ii} +  p_{\0} \sigma_{\ii} + \frac{m_{\ii}^{\2}}{2E_{\ii}} \sigma_{\ii}^{\2}.
\label{12}
\end{eqnarray}\normalsize
and get an analytic expression for the
oscillation probability.
The zero-order term in the previous expansion, $E_{\ii}$, gives the
standard plane-wave oscillation phase. The first-order term, $ p_{\0}
\sigma_{\ii}$, is responsible for the {\em slippage} between the mass-eigenstate wave packets
 due to their different group velocities. It represents a linear correction to
the standard oscillation phase \cite{DeL04}. Finally, the
second-order term, $\frac{m_{\ii}^2}{2E_{\ii}} \sigma_{\ii}^2$, which is a
(quadratic) secondary correction will give the well-known
{\em spreading} effects in the time propagation of the wave packet and
will be also responsible for an {\em additional} phase to be
computed in the final calculation.
In case of {\em gaussian}
momentum distributions, all these terms can be {\em analytically} quantified.
By substituting (\ref{12}) in Eq.~(\ref{4}) and changing the
$ p_{\z}$-integration into a $\sigma_{\ii}$-integration, we obtain the
explicit form of the mass-eigenstate wave packet time evolution,
\small\begin{eqnarray}
\phi_{\ii}(z,t)
 &=& 
 \left[\frac{2}{\pi \,a^{\2}_{\ii}(t)}\right]^{ \frac{1}{4}}\exp{[-i\,(\theta_{\ii}(t, z) + E_{\ii} \, t -  p_{\0} \, z)]}
 \exp{\left[-\frac{(z - \mbox{v}_{\ii} \,t)^{\2}}{a_{\ii}^{\2}(t)}\right]}
 ,~~~				  
\label{13}
\end{eqnarray}\normalsize
where
$\theta_{\ii}(t, z) = \left\{\frac{1}{2}\arctan{\left[\frac{2\,m_{\ii}^{\2}\, t}{a^{\2}\, E_{\ii}^{\3}}\right]} - \frac{2\, m_{\ii}^{\2}\, t} {a^{\2}\, E_{\ii}^{\3}}\,\frac{(z - \mbox{v}_{\ii} \,t)^{\2}}{a_{\ii}^{\2}(t)}\right\}$
and
$a_{\ii}(t) = a \left(1 + \frac{4\, m_{\ii}^{\4}}{a^{\4}\, E_{\ii}^{\6}}\,t^{\2}\right)^{ \frac{1}{2}}$.
The time-dependent quantities $a_{\ii}(t)$ and $\theta_{\ii}(t, z)$ contain all the physically significant information which arise from the second-order term in the power series expansion (\ref{12}).
By solving the integral (\ref{9}) with the approximation (\ref{11})
and performing some mathematical manipulations, we obtain
\small\begin{equation}
\mbox{\sc Int}(t) = \mbox{\sc Bnd}(t) \times \mbox{\sc Osc}(t),
\label{20}
\end{equation}\normalsize
where we have factored the time-vanishing bound of the interference term given by
\small\begin{equation}
\mbox{\sc Bnd}(t) = \left[1 + \mbox{\sc Sp}^{\2}(t) \right]^{-\frac{1}{4}}
\exp{\left[-\frac{(\Delta \mbox{v} \, t)^{\2}}{2a^{\2}\left[1 + \mbox{\sc Sp}^{\2}(t)\right]}\right]}
\label{21}
\end{equation}\normalsize
and the time-oscillating character of the flavor conversion formula given by
\small\begin{eqnarray}
\mbox{\sc Osc}(t) &=& Re \left\{\exp{\left[-i\Delta E \, t -i \Theta(t)\right]} \right\}\nonumber\\
&=& \cos{\left[\Delta E \, t + \Theta(t)\right]},
\label{22A}
\end{eqnarray}\normalsize
where
\small\begin{equation}
\mbox{\sc Sp}(t) = \frac{t}{a^{\2}}\Delta\left(\frac{m^{\2}}{E^{\3}}\right) = \rho\, \frac{\Delta \mbox{v}\, t}{a^{\2} \,  p_{\0}}
\label{240}
\end{equation}\normalsize
and
\small\begin{equation}
\Theta(t) = \mbox{$\left[\frac{1}{2}\arctan{\left[\mbox{\sc Sp}(t)\right]} - \frac{a^{\2} \,  p_{\0}^{\2}}{2 \rho^{\2}}\frac{\mbox{\sc Sp}^{\3}(t)}{\left[1 + \mbox{\sc Sp}^{\2}(t)\right]}\right]$}, 
\label{24A}
\end{equation}\normalsize
with
$\rho =
 1 - \left[3 + \left(\frac{\Delta E}{\bar{E}}\right)^{\2}\right] \frac{ p_{\0}^{\2}}{\bar{E}^{\2}}$, $\Delta E = E^{(\1)} - E^{(\2)}$ and $\bar{E} = \sqrt{E_{\1} \, E_{\2}}$.
The time-dependent quantities $\mbox{\sc Sp}(t)$ and $\Theta(t)$
carry the second-order corrections and, consequently, the
{\em spreading} effect to the oscillation probability formula.
If $\Delta E \ll \bar{E}$, the parameter $\rho$ is limited by
the interval $[1,-2]$ and it assumes the zero value when $\frac{ p_{\0}^{\2}}{\bar{E}^{\2}} \approx \frac{1}{3}$.
Therefore, by considering increasing values of $ p_{\0}$,
from non-relativistic (NR) to ultra-relativistic (UR) propagation regimes,
and fixing $\frac{\Delta E}{a^{\2} \, \bar{E}^{\2}}$,
the time derivatives of $\mbox{\sc Sp}(t)$ and $\Theta(t)$ have their
signals inverted when $\frac{ p_{\0}^{\2}}{\bar{E}^{\2}}$ reaches the value $\frac{1}{3}$.
The {\em slippage} between the mass-eigenstate wave packets is
quantified by the vanishing behavior of $\mbox{\sc Bnd}(t)$. 
In order to compare $\mbox{\sc Bnd}(t)$ with the correspondent
function without the second-order corrections (without {\em spreading}),
\small\begin{equation}
\mbox{\sc Bnd}_{\mbox{\tiny $WS$}}(t) = \exp{\left[-\frac{(\Delta \mbox{v} \, t)^{\2}}{2a^{\2}}\right]},
\label{23A}
\end{equation}\normalsize
we substitute ${\mbox{\sc Sp}(t)}$ given by the expression (\ref{22A})
in Eq.~(\ref{21}) and we obtain the ratio
\small\begin{eqnarray}
\frac{\mbox{\sc Bnd}(t)}{\mbox{\sc Bnd}_{\mbox{\tiny $WS$}}(t)}
&=& \mbox{$\left[1 + \rho^{\2} \left(\frac{\Delta E \, t}{a^{\2} \, \bar{E}^{\2}}\right)^{\2} \right]^{-\frac{1}{4}}$}
\mbox{$ \exp{\left[\frac{\rho^{\2} \,  p_{\0}^{\2} \,
\left(\Delta E \, t\right)^{\4}}
{2\,a^{\6} \, \bar{E}^{\8}\left[1 + \rho^{\2} \left(\frac{\Delta E \, t}{a^{\2} \, \bar{E}^{\2}}\right)^{\2}\right]}
\right]}.$}
\label{23AA}
\end{eqnarray}\normalsize
The NR limit is obtained by setting $\rho^{\2} = 1$ and $ p_{\0} = 0$ in Eq.~(\ref{23A}).
In the same way, the UR limit is obtained by setting $\rho^{\2} = 4$
and $ p_{\0} = \bar{E}$.
In fact, the minimal influence due to second-order corrections occurs
when $\frac{ p_{\0}^{\2}}{\bar{E}^{\2}} \approx \frac{1}{3}$ ($\rho \approx 0$).

\begin{figure}
\begin{center}
\hspace{-3cm}
\epsfig{file= 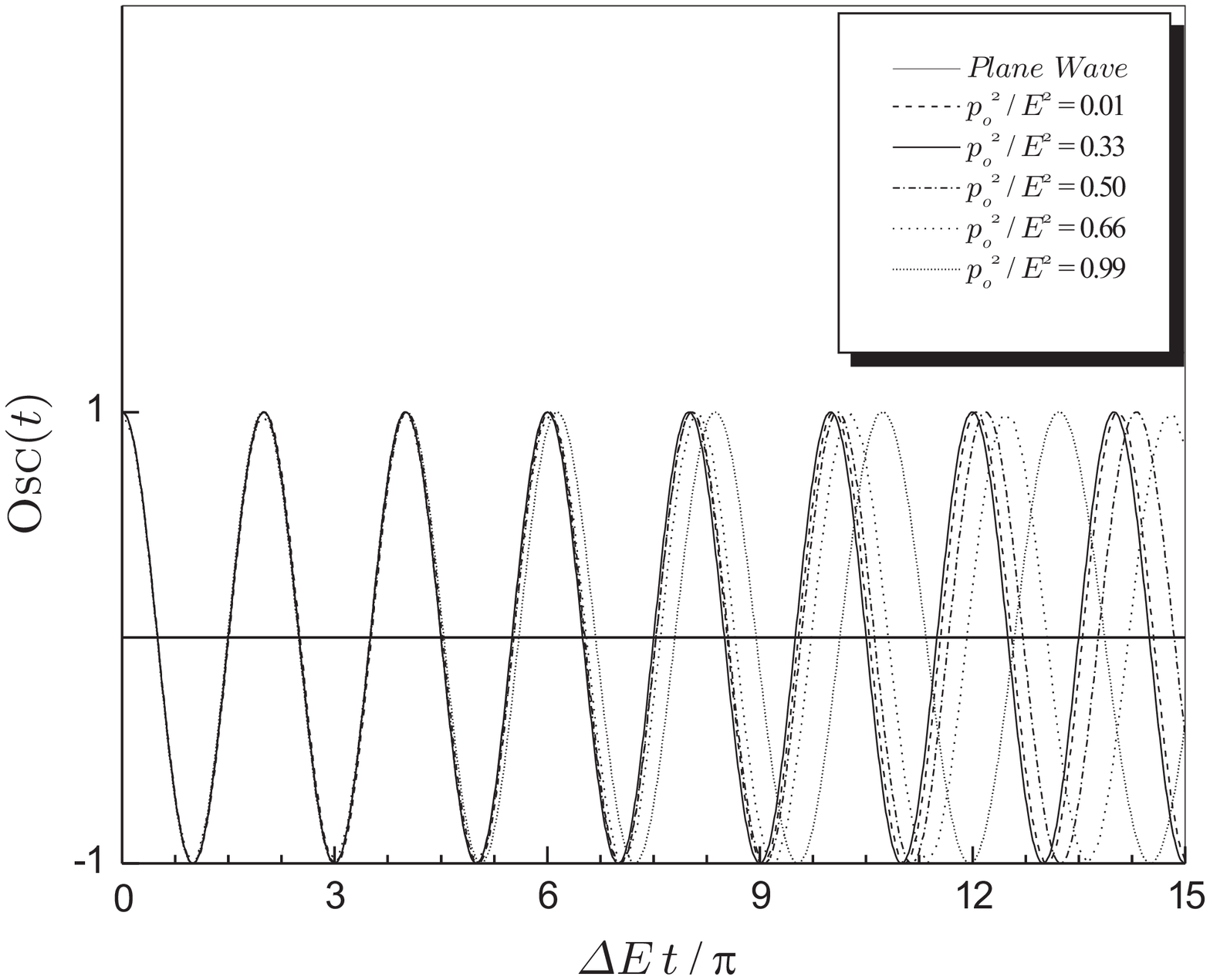, height= 15.0 cm, width= 19 cm}
\end{center}
\caption{\label{an4} The time-behavior of $\mbox{\sc Osc}(t)$ compared with the {\em standard} plane-wave oscillation given by $\cos{[\Delta E \, t]}$
for different propagation regimes.
The additional phase $\Theta(t)$ changes the oscillating character after some time of propagation.
The minimal deviation occurs for $\frac{p_{\0}^{\2}}{\bar{E}^{\2}} \approx \frac{1}{3}$ which is represented by a solid
line superposing the plane-wave case.
We have used $a \, \bar{E} = 10$ for this plot which was taken from reference \cite{Ber05}.}
\end{figure}

\begin{figure}
\epsfig{file= 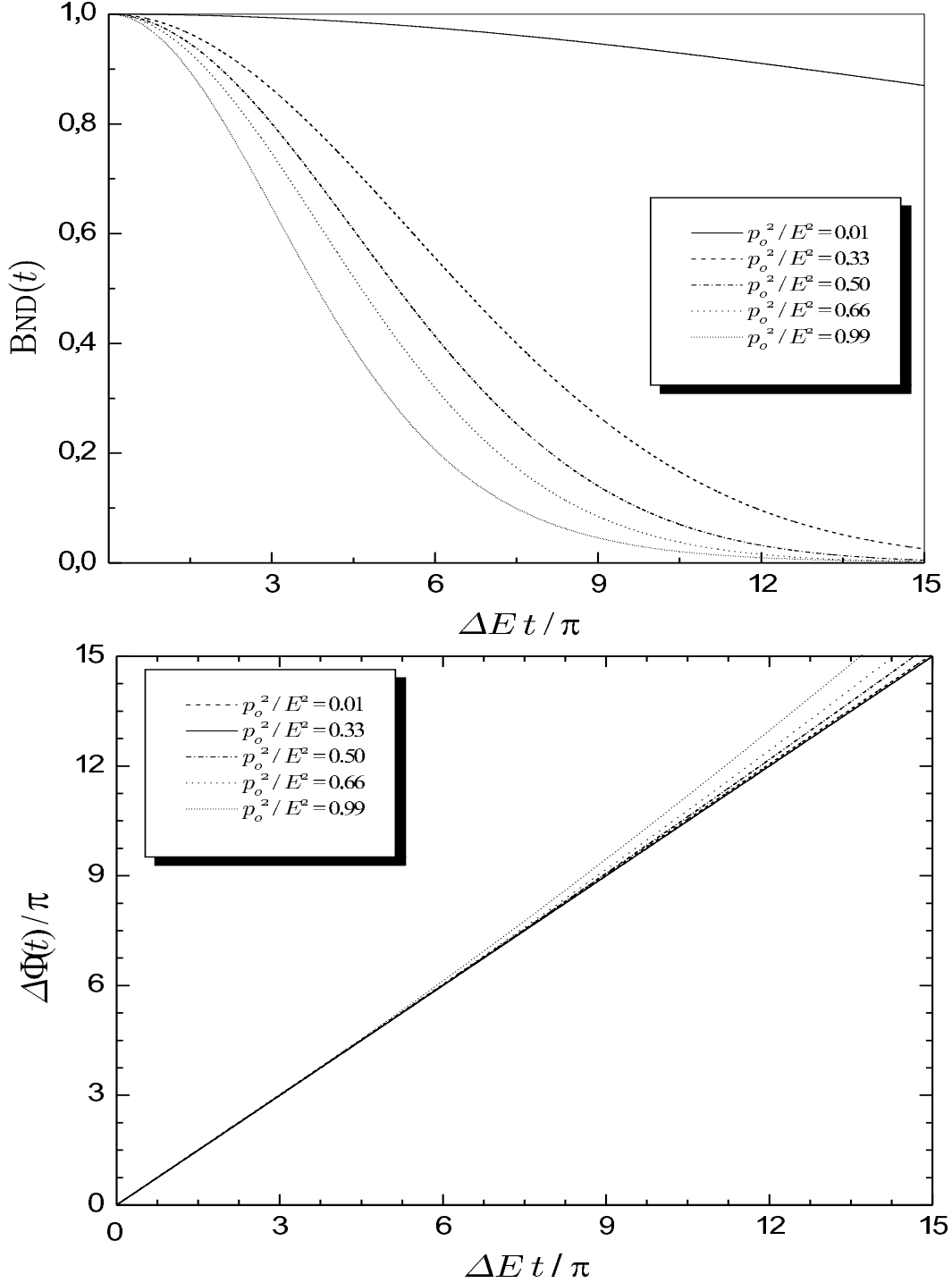, height= 19 cm, width= 14 cm}
\caption{\label{an5} We plot the behavior of the corrected phase $\Delta \Phi(t) = \Delta E \, t + \Theta(t)$ for different propagation regimes and we observe that the values assumed by $\Theta(t)$ are {\em  effective} only when the interference boundary function $\mbox{\sc Bnd}(t)$ does not vanish.
By diminishing the value of the wave packet parameter $a \, \bar{E}$ (we also have used $a \, \bar{E} = 10$ for this plot) the amortizing behavior is attenuated and the range of modifications introduced by the additional phase $\Theta(t)$ increases.
This plot was taken from reference \cite{Ber05}.}
\end{figure}

\begin{figure}
\hspace{-3cm}
\epsfig{file= 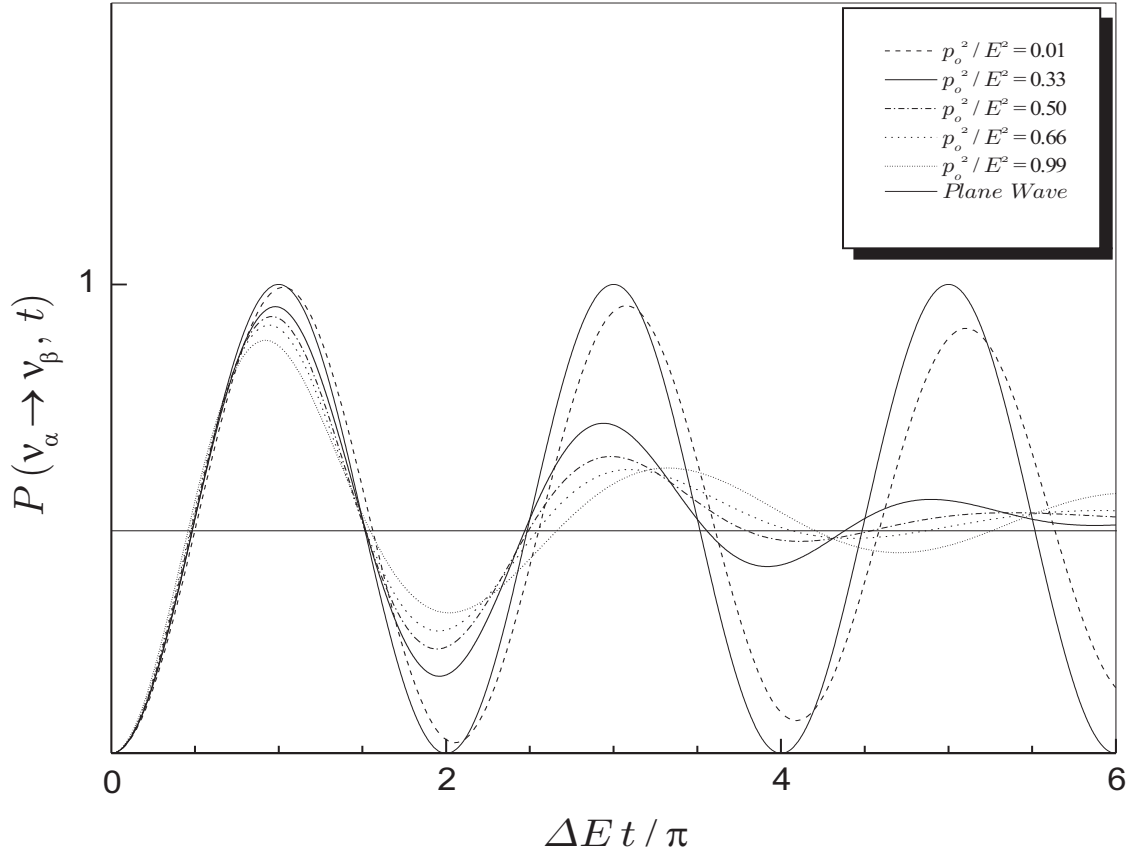, height= 15.0 cm, width= 19 cm}
\caption{\label{an8} The time-dependence of the flavor conversion probability obtained with the introduction of
second-order corrections in the series expansion of the energy for a strictly peaked momentum distribution ($\mathcal{O}(\sigma_{\ii}^{\3})$).
By comparing with the PW predictions, depending on the propagation regime,
the additional time-dependent phase $\Delta \Phi(t) \equiv\, \Delta E \, t + \Theta(t)$ produces a delay/advance in the local maxima of flavor detection.
Phenomenologically, we shall demonstrate that such modifications allow us to quantify small corrections to the averaged values of neutrino oscillation parameters, i. e. the mixing-angle and the mass-squared difference.
Essentially, it depends on the product of the wave packet width $a$ by the averaged energy $\bar{E}$.
Here again  we have used $a\bar{E} = 10$.}
\end{figure}

The oscillating function $\mbox{\sc Osc}(t)$ of the interference
term $\mbox{\sc Int}(t)$ differs from the {\em standard} oscillating
term, $ \cos{[\Delta E \, t]}$,
by the presence of the additional phase $\Theta(t)$
which is essentially a second-order correction.
The modifications introduced by the additional phase $\Theta(t)$ are discussed in Fig.~\ref{an4}
where we have compared the time-behavior of $\mbox{\sc Osc}(t)$ with $\cos{[\Delta E \, t]}$ for different propagation regimes.
The {\em  effective} bound value assumed by $\Theta (t)$
is determined by the vanishing behavior of $\mbox{\sc Bnd}(t)$.
To illustrate this flavor oscillation behavior, we plot both the curves representing $\mbox{\sc Bnd}(t)$ and $\Theta(t)$ in Fig.~\ref{an5}.
(both figures are obtained from \cite{Ber05}).
We note the phase slowly changing in the NR regime.
The modulus of the phase $|\Theta(t)|$ rapidly reaches its upper limit when $\frac{ p_{\0}^{\2}}{\bar{E}^{\2}} > \frac{1}{3}$ and, after a certain time, it continues to evolve approximately linearly in time.
Essentially, the oscillations vanishes rapidly.
By superposing the effects of $\mbox{\sc Bnd}(t)$ in Fig.~\ref{an5} and the oscillating character $\mbox{\sc Osc}(t)$ expressed in Fig.~\ref{an4}, we immediately obtain the flavor oscillation probability which is explicitly given by 
\small\begin{eqnarray}
P(\mbox{\boldmath$\nu_\alpha$}\rightarrow\mbox{\boldmath$\nu_\beta$};t)
 \approx
\frac{\sin^{\2}{(2\theta)}}{2}\left\{1 - \left[1 + \mbox{\sc Sp}^{\2}(t) \right]^{-\frac{1}{4}}
\exp{\left[-\frac{(\Delta \mbox{v} \, t)^{\2}}{2a^{\2}\left[1 + \mbox{\sc Sp}^{\2}(t)\right]}\right]}
\cos{\left[\Delta E \, t + \Theta(t)\right]}
  \right\}
\label{25A}
\end{eqnarray}\normalsize
and illustrated in Fig.~\ref{an8}.
Obviously, the larger is the value of $a \, \bar{E}$, the smaller are the wave packet effects.

\section{Understanding two-flavor oscillation parameters and limits}

As earlier extensively discussed in the literature \cite{Beu03,Kay81,Ric93,Giu98},
there are various restrictive conditions under which the two-neutrino mixing
approximation is valid and the standard PW as well as several classes of WP
treatment can be utilized for describing the flavor conversion phenomena.
Respecting such limitations, in the previous section we have quantified the
second-order corrections introduced by an intermediate WP analysis.
Now, we are in position to verify how such modifications can affect the reading of
the neutrino oscillation parameter ranges excluded by the confront with experimental
data.  
Therefore, our effective purpose is contrasting the standard PW treatment with the
WP analytic study presented in the previous section by re-obtaining the two-flavor
neutrino oscillation parameters and limits
in a very particular phenomenological context.

From a practical point of view, we have to establish the {\em input} experimental
parameters as being the detection distance $L_{\0}$ (from source), the neutrino energy
distribution $\bar{E}$ and the appearance (disappearance) probability $\langle P \rangle$.
In addition, to make clear the initial proposition, it is instructive to redefine
some parameters which shall carry the main physical information
in the oscillation formula, i. e.
\small\begin{equation}
b_{\0} = \frac{1}{2 \upsilon}\frac{L_{\0}}{E_{\1}+{E_{\2}}},
~~~~ \delta_{b} = \frac{b_{\0}}{a \, \bar{E}}
~~~~ \mbox{and}~~~~ \upsilon = \frac{p_{\0}}{\bar{E}},
\label{pap1}
\end{equation}\normalsize
with $\bar{E}$ previously defined.
For real experiments, $\bar{E}$ and $L_{\0}$ can have some spread due to various
effects, but in a subset of these experiments, there is a well-defined value of $b_{\0}$
about which the events distribute \cite{Gro04}.
Following the same approach we have adopted while we were analyzing
the parameter $\rho$ in Eq.~(\ref{240}), if  $\Delta E \ll \bar{E}$, which is perfectly acceptable from the
experimental point of view, we can write $\bar{E} = \sqrt{E_{\1} {E_{\2}}}
\approx \frac{1}{2}(E_{\1}+{E_{\2}})$ so that an effective PW flavor conversion
formula can be obtained from Eq.~(\ref{00}) as
\small\begin{equation}
\langle P \rangle_{\mbox{\tiny PW}} \equiv P(\mbox{\boldmath$\nu_\alpha$}
\rightarrow\mbox{\boldmath$\nu_\beta$};b_{\0}) = \frac{\sin^{\2}{(2\theta)}}{2} \left\{1 - \cos{[2 \, b_{\0} \, \Delta m^{\2}]}\right\}
\label{pap2}.
\end{equation}\normalsize
where $b_{\0}$ carries the dependence on the detection distance $L_{\0}$ and on the propagation regime ($\upsilon$).
Performing some analogous substitutions, the interference terms of Eq.~(\ref{20}) which explicitly appears in the
WP flavor conversion formula of Eq.~(\ref{25A}) can be read in terms of the above rewritten parameters
\small\begin{eqnarray}
\mbox{\sc Bnd}(b_{\0}) &=& \left[1 + \mbox{\sc Sp}^{\2}(b_{\0}) \right]^{\mi\frac{1}{4}}\exp{\left[- \frac{2 \delta_{b}^{\2}\,\upsilon^{\2}\, (\Delta m^{\2})^{\2}}{1 + \mbox{\sc Sp}^{\2}(b_{\0})}\right]},
~~~\label{pap3}
\end{eqnarray}\normalsize
and 
\small\begin{eqnarray}
\mbox{\sc Osc}(b_{\0}) &=& \cos{\left[2 \, b_{\0} \, \Delta m^{\2} + \Theta(b_{\0})\right]},
\label{pap4}
\end{eqnarray}\normalsize
where
\small\begin{equation}
\mbox{\sc Sp}(b_{\0}) = - \rho \frac{\delta_{b}^{\2}}{b_{\0}}\, \Delta m^{\2}
\label{pap5}
\end{equation}\normalsize
and
\small\begin{equation}
\Theta(b_{\0}) = \left[\frac{1}{2}\arctan{\left[\mbox{\sc Sp}(b_{\0})\right]} 
- \frac{b_{\0}^{\2} \,  \upsilon^{\2}}{2 \delta_{b}^{\2}\, \rho^{\2}}
\frac{\mbox{\sc Sp}^{\3}(b_{\0})}{\left[1 + \mbox{\sc Sp}^{\2}(b_{\0})\right]}\right], 
\label{pap6}
\end{equation}\normalsize
with $\rho \approx 1 - 3 \upsilon^{\2}$.
Attempting to the rate $\sigma = \delta_{b}/b_{\0} = (a \, \bar{E})^{\mi 1}$ which carries
the relevant information concerning with the wave packet width and the averaged energy flux,
if it is was sufficiently small ($\sigma \ll 1$) so that we could ignore
the second-order corrections of Eq.~(\ref{11}), the probability only with the leading terms
could be read as 
\small\begin{equation}
\langle P \rangle_{\mbox{\tiny WP1}} = \frac{\sin^{\2}{(2\theta)}}{2} 
\left\{1-\cos{[2  b_{\0} \, \Delta m^{\2}]}\,
\exp{\left[- 2 (\delta_{b}\,\upsilon\, \Delta m^{\2})^{\2}\right]}\right\},\label{pap7}
\end{equation}\normalsize
which brings up the idea of a {\em coherence length} $L_{coh} \sim \frac{a \bar{E}^{\2}}{\Delta m^{\2}_{\ii\j}}$ \cite{Beu03,Kie96}
and, in the particular case of an UR propagation ($\upsilon = 1$),
is taken as a reference in the confront with experimental data \cite{Gro04}.
By the way, despite the relevant dependence on the propagation regime ($\upsilon$),
once we are interested in some realistic physical situations,
the following analysis will be limited to the UR propagation regime corresponding to
the effective neutrino energy of the current flavor oscillation experiments.

\begin{figure}
\begin{center}
\epsfig{file= 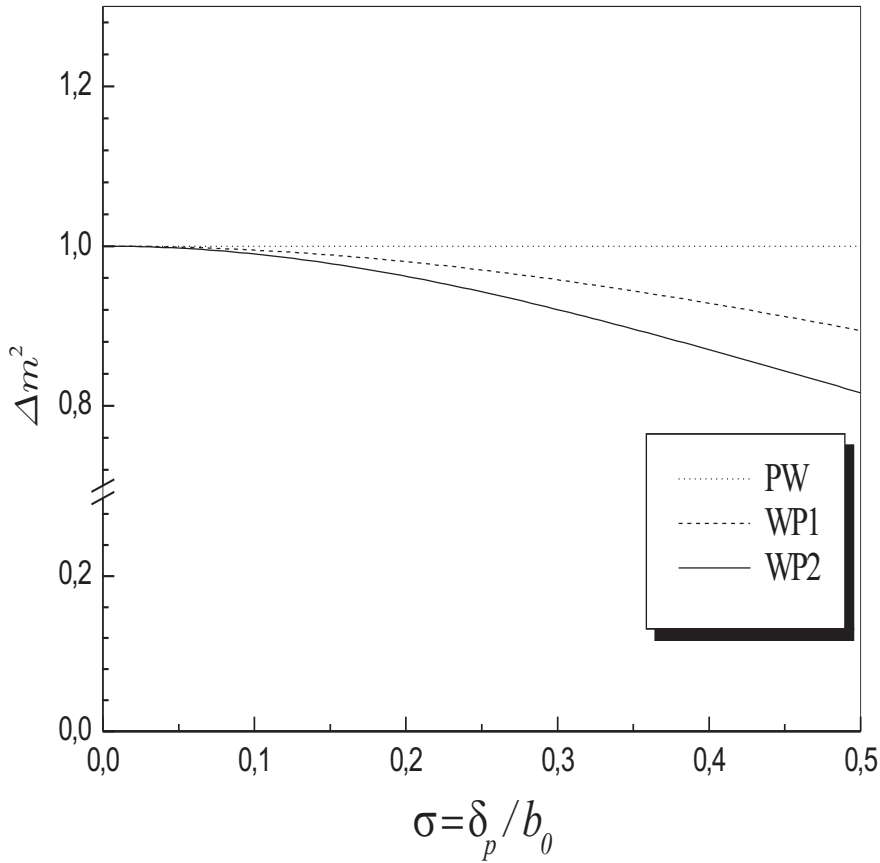, height= 15 cm, width=13.2 cm}
\vspace{-2cm}
\end{center}
\caption{\label{an9} Neutrino oscillation parameter ranges excluded by two experiments:
the Palo Verde reactor disappearance experiment (BOEHM 01) and the positive signal from the KamLAND collaboration (EGUCHI 03).
We process the input parameters by following the WP treatment with first and second-order corrections (WP1 and WP2)
where we have assumed the wave packet parameter $\sigma \sim 0.342$ which leads to a $10\%$ correction to the value of $\Delta m^{\2}$ obtained with the PW approximation.
In order to compare the results obtained with the WP approach with the PW prediction for the maximum excursion of the curve to the left,
which corresponds to an important phenomenological boundary region,
we have plotted the PW curve (dotted line) only for the smaller values of $\Delta m^{\2}$ (since it has to be averaged on the energy
for larger values).
The parameter ranges simultaneously allowed by both experiments are represented by the {\em not} filled area.
The second-order corrections introduce {\em accurate} modifications to the oscillation parameter ranges.
The nearer is $\sigma$ to 1, the more relevant is the contribution due to higher order terms in the Eq.~(\ref{12}) for determining such accurate limits.} 
\end{figure}

Strictly speaking, most results in the neutrino mixing listings are presented as $\Delta m^{\2}$ limits (or ranges)
for $\sin^{\2}(2\theta) = 1$ , and $\sin^{\2}(2\theta)$ limits (or ranges)
for large $\Delta m^{\2}$.
Together, they summarize the most of the information contained in the usual $\Delta m^{\2} \times \sin^{\2}(2\theta)$
plots which provide the parameter exclusion region boundary in the experiments' papers. 
Thus, we can compare the PW and WP resolutions by 
enumerating some relevant aspects which can be observed from
the curve $\Delta m^{\2} \times \sin^{\2}{(2\theta)}$ in the Fig.~\ref{an9} where 
the Palo Verde \cite{Boe01} and the KamLAND \cite{Egu03} experiments/exclusion curves are represented.
For the moment, we are interested in the analytical characteristics of such exclusion curves, which, however, we turn to analyze
under the phenomenological point of view in section IV.
The main aspects to be quantified are:

1) In both PW and WP cases, for large $\Delta m^{\2}$ the fast oscillations are
completely washed out respectively by the plane-wave resolution 
or by the smearing out behavior due to the mass-eigenstate wave packet decoherence. 
Consequently $\sin^{\2}{(2\theta)} = 2 \langle P \rangle$ in this limit.

2) For PW calculations the maximum excursion of the curve to the left occurs
at $\Delta m^{\2} = \pi / 2 b_{\0}$ when $\sin^{\2}{(2\theta)} = \langle P \rangle$.
When the WP Eq.~(\ref{pap7}) is used such a maximal point occurs at 
the solution of the transcendental equation $-2\sigma^{\2} \,\Delta m^{\2}\, b_{\0} =\tan{[2 \,\Delta m^{\2}\, b_{\0}]}$ 
which can be approximately given by $\Delta m^{\2} \approx \pi / 2\, b_{\0}\left(1+ \mathcal{O}(\sigma^{\2})\right)$,
but we know that the second-order terms ($\sigma^{\2}$) are not being considered. If we had taken into account the second-order corrections
in the WP analysis, the maximal value would have been accurately given by
$\Delta m^{\2} \approx \pi / 2\, b_{\0}\left(1+ 2\sigma^{\2} + \mathcal{O}(\sigma^{\4})\right)$,
consequently, a little smaller value than the PW solution.

\begin{figure}
\epsfig{file= 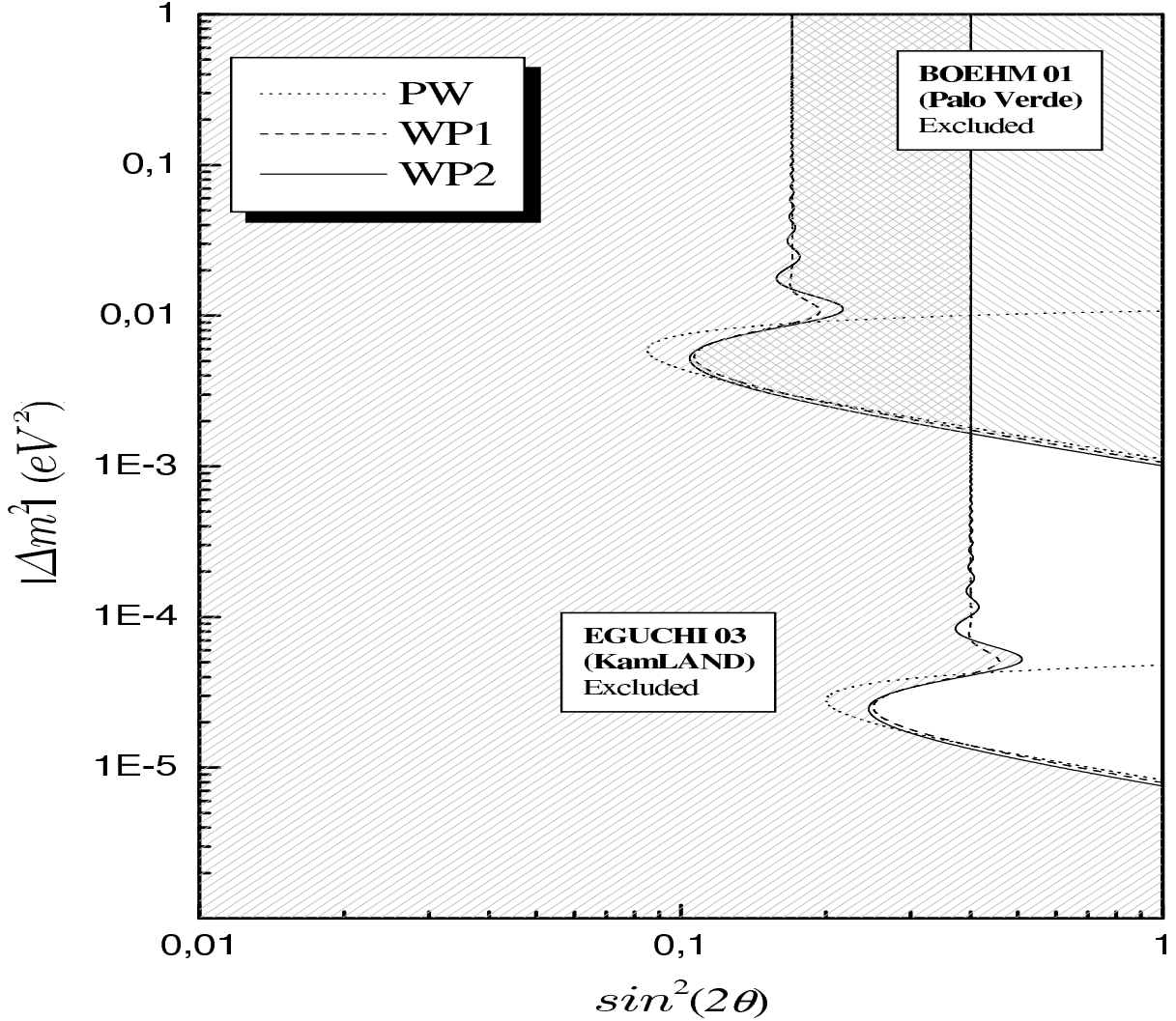, height= 14 cm, width= 16 cm}
\caption{\label{an10} Relative corrections to the $\Delta m^{\2}$ value (obtained with the PW analysis)
by assuming the WP treatments with first (WP1) and second-order (WP2) corrections in the energy expansion.
We normalize the result by dividing $\Delta m^{\2}$ by
$\frac{\sqrt{\langle P \rangle }}{b_{\0} \, \sin{(2\theta)}}$
where obviously $\langle P \rangle _{\mbox{\tiny PW}} = \langle P \rangle _{\mbox{\tiny WP1}} = \langle P \rangle _{\mbox{\tiny WP2}} = \langle P \rangle$
is obtained from the experimental results.
For instance, by following the WP treatment with second-order corrections,
the correction to the phenomenological parameter $\Delta m^{\2}$ corresponds
to a diminution of approximately $10\%$ of the $\Delta m^{\2}$
PW value when $\sigma \approx 0.342\, (a\bar{E}\approx 3)$, 
$1\%$ when $\sigma \approx 0.100 \, (a\bar{E}\approx 10)$,
and  $0.1\%$ when $\sigma \approx 0.032\, (a\bar{E}\approx 30)$. } 
\end{figure}

3) By qualitatively assuming the well-established phenomenological constraints which set $\Delta m^{\2} \ll 1 \,eV^{\2}$
when $\sin^{\2}(2\theta) \approx 1$, we can reconstruct the nearly straight-line segment at the bottom of the curve
by expanding the probability expressions up to order $\mathcal{O}((\Delta m^{\2})^{\2})$ so that we
can obtain the generic solution 
\small\begin{equation}
\Delta m^{\2} \approx \frac{\sqrt{\langle P \rangle _{\mbox{\tiny App}}}}{b_{\0} \, \sin{(2\theta)} \,\sqrt{F_{\mbox{\tiny App}}(\sigma)}}
\label{pap8}
\end{equation}\normalsize
where we have $F_{\mbox{\tiny PW}}(\sigma) = 1$ for the PW limit, $F_{\mbox{\tiny WP1}}(\sigma) = 1 +  \sigma^{\2}$ for the WP treatment
with first-order corrections and $F_{\mbox{\tiny WP2}}(\sigma) = 1 + 2\sigma^{\2} + \frac{3}{4}\sigma^{\4}$ for the WP treatment
with second-order corrections.
Eventually, if one had abandoned the analytic calculations and had
taken into account higher order terms in the power series expansion of Eq.~(\ref{12}),
there would have been some minor corrections to the $\sigma^{\4}$ term in $F_{\mbox{\tiny WP2}}$.
We discard such minor corrections by assuming $F_{\mbox{\tiny WP2}}(\sigma)\approx 1 + 2 \sigma^{\2}$
which, in fact, is the correct approximation when we are considering the energy expansion up to second-order terms
and sufficient for comparing the approximations in the Fig.~\ref{an10}.
We emphasize that we {\em must} consider the second-order term in the series expansion in the Eq.~(\ref{12})
since the modifications emerge with $\sigma^{\2}$.
As a {\em testing} case, we consider the {\em toy} model presented in \cite{Gro04} where the values of $\sigma = 0.23$ and $\sigma = 0.3$ are
used for plotting the curve $\Delta m^{\2} \times \sin^{\2}{(2\theta)}$.
From an immediate analysis of the  Eq.~(\ref{pap8}) with WP2 approximation we can conclude that
more accurate values of $\Delta m^{\2}$ are constrict to be diminished by approximately $8 \%$
of the value computed with the PW approximation.

Quite generally, the complete analysis of the amortisation coupled to the oscillating character depends upon the experimental features such as the size of the source,
which allows us estimating the wave packet width ($a$), the neutrino energy distribution ($\bar{E}$) and the detector resolution ($L_{\0}$).
Anyway, the main point to be considered is that the characterization of the wave packed ($a$) implicitly described by $\sigma$, which is accompanied by the precise determination of the neutrino energy distribution ($\bar{E}$), plays a fundamental role when the neutrino oscillation parameter and limit accuracy is the subject of the phenomenological analysis.
To clarify this point, in addition to the information interpreted from Fig.~\ref{an10}, in the next section we turn back to the phenomenological discussion on the Fig.~\ref{an9}
which illustrates one simple case of parameter determination for reactor neutrinos by considering the results from some {\em disappearance} experiments \cite{Egu03,Boe01}.

\section{Phenomenological constraints to reactor experiments - Extension to solar and supernova neutrinos}

The first hints that neutrino oscillations actually occur were serendipitously obtained through early studies of solar neutrinos \cite{Fuk02} and neutrinos produced in the atmosphere by cosmic rays \cite{Fuk00,Amb98}.
More recently, nuclear reactors and particle accelerators have constituted another source of neutrinos utilized for accurate measurements of flavor oscillation parameters and limits.
Reactor neutrino experiments correspond essentially to an electron-antineutrino $\bar{\nu}_{e}$ {\em disappearance} experiment where,
generically speaking, one looks for the attenuation of the initial neutrino flavor-eigenstate $\mbox{\boldmath$\nu_\alpha$}$ beam in transit to a detector, where the $\mbox{\boldmath$\nu_\alpha$}$ is measured.
In contrast to the detection of even a few {\em wrong-flavor} ($\mbox{\boldmath$\nu_\beta$}$) neutrinos establishing mixing in an {\em appearance} experiment, the disappearance of a few {\em right-flavor} ($\mbox{\boldmath$\nu_\alpha$}$) neutrinos goes unobserved because of statistical fluctuations \cite{Gro04}.
For this reason, disappearance experiments usually cannot establish small-probabilities ($\sin^{\2}{(2\theta)} \ll 1$).
Besides, they can fall into several situations \cite{Gro04} into which we do not intend to go deep.

By following the purpose of a comparative phenomenological study, we take into account the experimental data from the Palo Verde \cite{Boe01} and KamLAND \cite{Egu03} experiments by means of which exclusion plots in the plane $\Delta m^{\2} \times \sin^{\2}{(2\theta)}$ can be elaborated.
The Palo Verde experiment \cite{Boe01} consists in a disappearance search for $\bar{\nu}_{e}$ oscillations at $0.75-0.89\,km$ distance from the Palo Verde reactors.
As consequence of the experimental analysis we assume the input parameter $\langle P \rangle < 0.084$ and the averaged energy $\bar{E}$ set in the interval $3.5-4.2\, MeV$.
The KamLAND collaboration \cite{Egu03} observes reactor $\bar{\nu}_{e}$ disappearance at $\sim 180\, km$ baseline to various Japanese nuclear power reactors.
This is the lower limit on the mass difference spread unlike all other disappearance experiments \cite{Egu03} and the observation is consistent with neutrino oscillations, with mass-difference and mixing angle parameters in the Large Mixing Angle Solution region of the solar neutrino problem.
In this case, we assume the input parameter $\langle P \rangle > 0.2$ with the reactor $\bar{\nu}_{e}$ energy spectrum smaller than $8 \, MeV$ and an analysis threshold of $2.6 \, MeV$ where the experiment sensitive $\Delta m^{\2}$ range is set down to $10^{\mi \5} eV^{\2}$ \cite{Egu03}.
The important point we attempt in the Fig.~\ref{an9} is the comparative modifications due to second-order corrections, i. e. we are not setting extremely accurate input (experimental) parameters but we are setting a more accurate procedure for obtaining the output parameter ($\Delta m^{\2}$ and $\theta$) ranges and limits.

The analysis of solar neutrino \cite{Fuk02,Ahm02} measurements involves considerable input from solar physics and the nuclear physics involved in the extensive chain of reactions that together are termed ``Standard Solar Model'' (SSM) \cite{SSM}. 
Since the predicted flux of solar neutrinos from SSM is very well-established \cite{SSM} we know that the low energy $p-p$ neutrinos are the most abundant and, since they arise from reactions that are responsible for most of the energy output of the sun, the predicted flux of these neutrinos is constrained very precisely ($\pm 2 \%$) by the solar luminosity.
The same is not true for higher energy neutrinos for which the flux is less certain due to uncertainties in the nuclear and solar physics required to compute them.

In fact, the true frontier for solar neutrino experiments is the real-time, spectral measurement of the flux of neutrinos below $0.4 \, MeV$ produced by $p-p$ reactions.
Measurements of the $p-p$ flux to an accuracy comparable to the accuracy of the SSM calculation will significantly improve the
precision of the mixing angle \cite{Bah03,Ban03}.
In addition, if the total $p-p$ flux is well-know, measurement of the active component can help constrain a possible sterile admixture \cite{Gro04}.
Therefore, an accurate phenomenological analysis for obtaining small modifications to the oscillation parameters, as we have illustrated in the Fig.~\ref{an9}, can be really pertinent.
In this context, the experimental challenge is to achieve low background at low energy threshold.
In particular, a number of projects and proposals aiming to build $p-p$ neutrino
spectrometers are discussed in \cite{Der04} where it has been shown that a large volume liquid organic scintillator detector
with an energy resolution of $10\, keV$ at $200\, keV$ can be sensitive to solar $p-p$ neutrinos,
if operated at the target radiopurity levels for the Borexino detector, or the solar neutrino project of KamLAND \cite{Egu03}.
In spite of the higher energy neutrinos being more accessible experimentally, the corrections to the wave packet
formalism can be physically relevant for $p-p$ neutrinos with energy distributed around an averaged value of
$\bar{E} \approx 10-100 \,keV$.
Following some standard procedure \cite{Kim93} for calculating the neutrino flux wave packet width $a$ for $p-p$ solar reactions, we obtain $a = 10^{\mi 10}- 10^{\mi 8}\, m \equiv 0.5-50\, (keV)^{\mi 1}$.
Such an interval sets a very particular range for the $\sigma$ parameter comprised by the interval
$5\cdot 10^{\mi \5} - 0.2$ for $\bar{E} \approx 0.01-0.4\, MeV$ which introduces the possibility for
WP second-order corrections establish some not ignoble modifications for the $p-p$ neutrino oscillation parameter limits. 

Turning back to the confront between the PW and the WP formalism, once we had precise values for the input parameters $\langle P \rangle$, $L_{\0}$ and $\bar{E}$,we could determine the effectiveness of the first/second-order corrections in determining $\Delta m^{2}$ for any class of neutrino oscillation experiment.
For instance, the flux of atmospheric neutrinos produced by collisions of cosmic rays (which are mostly protons) with the upper atmosphere is measured by experiments prepared for observing $\nu_{e} \leftrightarrow \nu_{\mu}$ and $\bar{\nu}_{e} \leftrightarrow \bar{\nu}_{\mu}$ conversions.
In particular, SuperKamiokande \cite{Fuk00} and MACRO \cite{Amb98} measurements also work on the $\nu_{\mu} \leftrightarrow \nu_{\tau}$ conversion. 
The neutrino energies range about from $0.1\, GeV$ to $100\, GeV$ which constrains the relevance of WP effects to an wave packet width $a \sim 10^{\mi\1\2}\, m$.
The accelerators experiments \cite{Ana98,Agu01,Arm02} cover a higher variety of neutrino flavor conversions where the neutrino energy flux stays around $1 - 10 \, GeV$ the limitations to the wave packet width $a$ (see the Table \ref{tab1}) are analogous which makes the WP second-order corrections, as a first analysis, completely irrelevant.

The most prominent contribution from the above discussion in determining the oscillation parameters and limits
can come with the analysis of supernova neutrinos which, however, are not yet solidly established by the experimental data.
Neutrinos from SN1987A in the Large Magellanic Cloud were detected by Kamiokande and IMB detectors - only 19 events.
Nowadays, SuperKamiokande, SNO, LVD, ICARUS, IceCube are expected to detect events from the next galactic supernova
and improve the statistics, providing new information on neutrino properties and supernova.
The main problem in studying neutrinos oscillation from supernova is the spectral and temporal evolution of
the neutrino burst.
In a supernova, the size of the wave packet is determined by the region of production (plasma), due to a process known as pressure broadening, which depends on the temperature, the plasma density, the mean free path of the neutrino producing particle and its mean termal velocity \cite{Kim93}.
Neutrinos from supernova core with $100\,MeV$ energy have a wave packet size varying from
$\sim 5\cdot 10^{\mi \1 \6} \,m$ to $\sim 10^{\mi \1 \4} \,m$ which leads to a wave packet
parameter $a \bar{E}$ comprised by the interval $0.25 - 5$ for which the second-order corrections
can be relevant if $a \bar{E} > 1$.
There also are $10\,MeV$ neutrinos, from neutrinosphere, with a wave packet size of approximately $10^{\mi\1\2}\,m$ with $a \bar{E} \approx 50$.        

In table \ref{tab1} we summarize the results for the five well-known neutrino sources:
reactor, solar, atmospheric, accelerator and supernova.
We compare the size of the wave packet $a$ determined by the region within which the production process of a neutrino is effectively
localized by the physical process itself (for instance, from a cross section analysis) with the
limits below which the WP second-order corrections
set a more accurate value to the mass-squared difference $\Delta m^{\2}$.
It allow us to establish where/when the physical effects due to second-order corrections can be eventually be expected.

The limits presented in table \ref{tab1} represents, for instance, a diminution of $10\%$, $1\%$ and $0.1\%$ from the value computed with the PW approach.
However, a pertinent objection to the the representativeness of such values can be stated:
it is important to observe that neutrino energy measurements cannot be performed very precisely so that
this produces an effect competing with that of the finite size of the wave packet.
If we set the energy uncertainty represented by $\delta E$, the Heisenberg 
uncertainty relation states that $\delta E\, a \sim 1$ and, consequently, our approximation hypothesis leads to
$\frac{\delta E}{\bar{E}} \sim \frac{1}{a\,\bar{E}} \ll 1$.
Realistically speaking, a typical neutrino-oscillation experiment searches for flavor conversions 
by means of an apparatus which, apart from the details inherent to the physical process,
provides an indirect measurement of the neutrino energy (in each event) accompanied by an experimental error
$\Delta E_{exp}$ due to the ``detector resolution''.
In case of $\frac{\Delta E_{exp}}{\bar{E}}<\frac{\delta E}{\bar{E}} \sim \frac{1}{a\,\bar{E}}$
the effective role of the second-order corrections illustrated in this analysis can be relevant.
On the contrary, $\Delta E_{exp} > \delta E$ demands for an averaged energy integration where
the decoherence effect through imperfect neutrino energy measurements by far dominates.
In a quantitative analytical analysis, this problem could be overcome by performing an analytical energy integration which, in general, is not possible.
In the present context, the current experimental values/measurements set some limitations to the applicability of our analysis by restricting it
to the $^{\7}Be$ and $p e^{\mi}p$ lines for solar neutrinos for which the energy flux is precisely defined
($\frac{\Delta E_{exp}}{\bar{E}} \ll 1$) and, eventually, to some (next generation)
reactor experiment and, certainly, to supernova neutrinos.

As an additional remark, without comprehending the exact theoretical coupled to experimental procedure
for determining the wave packet widths for a certain type of neutrino flux, we cannot arbitrarily assume,
apart from the {\em obvious} criticisms to the PW approach, the modifications introduced by the WP treatment
(particularly with second-order corrections) are irrelevant in the confront with a generic class of neutrino experimental data.
Finally, from the phenomenological point of view, the general arguments presented in \cite{Gro04} continue to be valid, i. e.
the above discussion has so far been limited to {\em vacuum} oscillations, where the mixing probabilities are described in terms
of the mixing angle.
In the solar neutrino case it is likely that interactions between the neutrinos and solar electrons affect the oscillation probability \cite{MSW}.

\section{Conclusion}

We have reported about the intermediate WP prescription in the context of neutrino flavor oscillations
from which, under the particular assumption of a sharply peaked momentum distribution
which sets an analytical approximation of order $\mathcal{O}((a\bar{E})^{\mi \3})$,
we have re-obtained an explicit expression for the flavor conversion formula
for (U)R and NR propagation regimes.
By concentrating our arguments on quantifying the second-order effects of such an approximation,
we have observed that
the existence of an additional time-dependent phase in the
oscillating term of the flavor conversion formula
and the modified {\em spreading} effect can represent some minor but accurate
modifications to the (ultra)relativistic oscillation probability formula which leads to important
corrections to the phenomenological analysis for obtaining the neutrino oscillation parameter and limits.

In particular, we have quantified such effects for determining the corrections to the
$\Delta m^{\2} \times \sin^{\2}{(2\theta)}$ curve for two reactor experiments \cite{Boe01,Egu03}.
The oscillation parameter range deviation from the PW values depends effectively on the product between
the wave packet width $a$ and the averaged energy flux $\bar{E}$ which characterizes the detection process.
The importance of the second-order corrections which come from the WP construction can also be relevant
in the framework of three-neutrino mixing.
It is well diffused that the next question which can be approached experimentally is that of $e3$ mixing.
A consequence of a non-zero $U_{e\3}$ matrix element will be a small appearance of $\nu_{e}$ in a bean of
$\nu_{\mu}$: for the particular case where $\Delta m^{\2}_{\1\2} \ll \Delta m^{\2}_{\2\3}$ (experimental data),
and for $\bar{E}_{\nu} \sim L \,\Delta m^{\2}_{\2\3}$, ignoring matter effects, we can find \cite{Kim93}
\small\begin{equation}
P(\mbox{\boldmath$\nu_{\mu}$}\rightarrow\mbox{\boldmath$\nu_{e}$};L)=
\sin^{\2}{(2\theta{\1\3})}\,\sin^{\2}{(\theta{\2\3})}\sin^{\2}{\left[\frac{\Delta m^{\2}_{\2\3}\, L}{4 \bar{E}}\right]}.
\label{100}
\end{equation}\normalsize
This expression illustrates that $\theta_{\1\3}$ manifests itself in the amplitude of an oscillation
with 2-3 like parameters.
By assuming an intermediate wave packet analysis, {\em fine-tuning} corrections can eventually be relevant. 

Experimentally, since the modulation may be parts per thousand or smaller, one needs both good statistics and low background data.
For instance the KamLAND experiment will significantly reduce the allowed region for
$\Delta m^{\2}_{\1\2}$ and $\sin{(2\theta_{\1\2})}$ relative to the present results, where the second-order
wave packet corrections can appear as an additional ingredient for accurately applying the phenomenological analysis.
At the same time, the next major goal for the reactor neutrino program will be to attempt a measurement of $\sin^{\2}{(2\theta_{\1\3})}$.
It can be shown that the reactor experiments have the potential to determine $\theta_{\1\3}$ without ambiguity
from CP violation or matter effects (by assuming the necessary statistical precision which requires large reactor
power and large detector size).
With reasonable systematic errors ($< \,1 \%$) the sensitivity is supposed to reach about
$\sin^{\2}{(2\theta_{\1\3})}\approx 0.01-0.02$ \cite{Vog04} and an accurate method of analysis, maybe in the wave packet framework,
can be required.

Turning back to the foundations for applying the intermediate wave packet formalism in the
neutrino oscillation problem, we know the necessity of a more sophisticated approach
is required. It can involve a field-theoretical treatment.
Derivations of the oscillation formula resorting to field-theoretical methods are not very
popular. They are thought to be very complicated and the existing quantum field computations of the
oscillation formula do not agree in all respects \cite{Beu03}.
The  Blasone and Vitiello (BV) model \cite{Bla95,Bla03} to neutrino/particle mixing and oscillations
seems to be the most distinguished trying to this aim.
They have attempt to define a Fock space of weak eigenstates to derive a nonperturbative oscillation formula.
Also with Dirac wave packets, the flavor conversion formula can be reproduced \cite{Ber04B}
with the same mathematical structure as those obtained in the BV model \cite{Bla95,Bla03}.
In fact, both frameworks deserve a more careful investigation since
the numerous conceptual difficulties hidden in the quantum oscillation phenomena still
represent an intriguing challenge for physicists.

\begin{acknowledgments}
This work was supported by FAPESP (04/13770-0).
\end{acknowledgments}

\newpage

\begin{table*}
\caption{\label{tab1} Wave packet width $a$ estimative for which the WP second-order corrections start to be relevant
in obtaining an accurate value of $\Delta m^{\2}$.
We have computed the $a$ limit by assuming $10\%(\sigma \sim 0.342)$, $1\%(\sigma \sim 0.100)$ and $0.1\%(\sigma \sim 0.032)$,
corrections to the value computed with the PW approximation.}
\footnotesize
\begin{tabular}{l c|c|r|r|r|r|r}
\hline\hline
$\nu$ class & Process(Abbr.) & Ref. & $\bar{E}(MeV)$ &     $a_{\1} (10^{\mi \1\2}m)$~~   &     $a_{\2} (10^{\mi \1\2}m)$~~&     $a_{\3} (10^{\mi \1\2}m)$~~& $a_{\mbox{\tiny Theo.}}(m)$\\
            &                &      &                &  ($10\%$ )& ($1\%$)& ($0.1\%$)&Ref.\cite{Kim93}\\ 
\hline\hline
Reactor     &$\nu_{e}\rightarrow\hspace{-0.35cm}\backslash\hspace{0.35cm}\nu_{e}$&\cite{Egu03}, \cite{Boe01}&$\sim 3$			 &$\sim 0.2$&$\sim 0.7$&$\sim 2.1 $ & $\sim 10^{\mi\6}$ \\
\hline
	        &$pp$                                                          &  \cite{SSM}            & $\sim 0.1$         &$\sim 6$ &$\sim 20.0$&$\sim 63.2  $&\\
	        &$^{\7}Be$                                                     &  \cite{SSM}			& $\sim 0.9$         &$\sim 1.5$ &$\sim 5.0$&$\sim 15.8  $&\\
Solar       &$^{\1\3}N,\,^{\1\5}O,\,^{\1\7}F$ 							   &  \cite{SSM}			& $\sim ~~1$         &$\sim 0.6$ &$\sim 2.0$&$\sim 6.3$& $\sim 10^{\mi\9}$\\
($BP2000$)  &$pe^{\mi}p$												   &  \cite{SSM}			& $\sim 1.4$         &$\sim 0.4$ &$\sim 1.4$&$\sim 4.5  $&\\
		    &$^{\8}B,\,hep$                                                &  \cite{SSM}			& $\sim ~10$         &$\sim 0.06$&$\sim 0.2$&$\sim 0.6$&\\
\hline
Atmospheric &$\nu_{\mu}\rightarrow\nu_{e,\tau, s}$                         & \cite{Fuk02,Amb98} & $10^{\2\,\mbox{\tiny to}\,\5}$  &$6\cdot 10^{\mi\3\,\mbox{\tiny to}\,\mi\6}$&$2\cdot 10^{\mi\2\,\mbox{\tiny to}\,\mi\5}$&$6\cdot 10^{\mi\2 \,\mbox{\tiny to}\,\mi\5}$& $\mi\mi\mi$\\
\hline
Accelerator &$\nu_{\mu}(\bar{\nu}_{\mu})\rightarrow\nu_{e}(\bar{\nu}_{e})$ & \cite{Agu01}       & $10^{\3\,\mbox{\tiny to}\,\4}$  &$6\cdot 10^{\mi\4\,\mbox{\tiny to}\,\mi\5}$&$2\cdot 10^{\mi\3 \,\mbox{\tiny to}\,\mi\4}$&$6\cdot 10^{\mi\3 \,\mbox{\tiny to}\,\mi\4}$& $\sim 1$\\
\hline
Supernova & \mi\mi\mi                                                           &   \mi\mi\mi              & $\sim ~10^{2}$            &$\sim 6\cdot 10^{\mi\3}$& $\sim 2\cdot 10^{\mi\2}$&$\sim 6\cdot 10^{\mi\2}$& $\sim 10^{\mi\15}$\\
(estimative)&&&&&&\\
\hline\hline
\end{tabular}
\end{table*}
\vspace{15cm}
\pagebreak

\end{document}